\documentclass[sigconf]{acmart}
\usepackage{graphicx}
\usepackage{booktabs}
\usepackage{xcolor}
\usepackage{listings}
\usepackage{subcaption}

\title{Syntax Is Not Enough: An Empirical Study of Small Transformer Models for Neural Code Repair}

\author{Shaunak Samant}
\affiliation{%
  \institution{MIT World Peace University}
  \city{Pune}
  \state{Maharashtra}
  \country{India}
}
\email{shaunaksam47@gmail.com}

\begin{abstract}
Automated program repair using neural models has shown promising results on benchmark datasets, yet practical deployment remains limited. In this study, we examine whether a small transformer model can meaningfully repair real-world Java bugs, and whether syntactic correctness is a reliable proxy for semantic correctness.

We fine-tune CodeT5-small (60.5M parameters) on 52,364 Java bug-fix pairs from CodeXGLUE and evaluate both token-level performance and syntactic validity using AST parsing. While the model converges cleanly and achieves high grammatical correctness—producing syntactically valid Java code in 94 percent of cases—it fails to generate correct repairs under exact-match evaluation, achieving no accuracy. In 80 percent of cases, the model reproduces the buggy input verbatim.

Through quantitative analysis and manual inspection, we show that high syntax validity does not imply repair capability. We attribute this failure to three factors: identifier abstraction that removes semantic signals, cross-entropy training objectives that reward conservative copying, and insufficient model capacity for multi-step program reasoning.

Our results suggest that commonly reported metrics may substantially overestimate the practical effectiveness of small neural repair models. We argue that future work should prioritize semantically informed datasets, execution-aware objectives, and evaluation methods that go beyond surface-level correctness.
\end{abstract}


\begin{document}
\setcopyright{none}
\acmConference{}{}{}
\acmBooktitle{}
\acmYear{}
\acmISBN{}
\acmDOI{}

\maketitle

\section{Introduction}

Software bugs impose massive economic costs—recent estimates place annual losses at over \$2 trillion globally~\cite{software-fails-2020}. While testing and debugging tools have improved, the core challenge remains: developers spend substantial time manually identifying and fixing defects. Automated program repair (APR) aims to reduce this burden by generating patches algorithmically. Early APR systems relied on predefined templates or constraint solvers, which worked well for specific bug classes but struggled to generalize.

The rise of deep learning has opened new possibilities. Large code corpora from GitHub and other repositories provide millions of examples showing how developers fix bugs in practice. Neural models can potentially learn these repair patterns through supervised training. Recent transformer-based approaches—including CodeBERT~\cite{codebert}, GraphCodeBERT~\cite{graphcodebert}, and CodeT5~\cite{codet5}—have achieved notable results on code completion, translation, and summarization tasks. This success naturally raises the question: can these models learn to repair code?

A key practical concern emerges: generated repairs must compile before they can be tested. Prior work observes that 20-40\% of neural model outputs contain syntax errors~\cite{code-repair-syntax}, rendering them useless regardless of their semantic intent. This motivated our investigation into whether smaller, efficient models can produce grammatically correct repairs, and whether grammatical correctness predicts actual bug fixes.

We focus on CodeT5-small, a 60.5M parameter model that fits on consumer GPUs. Training larger models is expensive and often impractical for individual researchers or small teams. If compact models can achieve high syntax validity, they become viable tools. However, our results reveal a troubling pattern: the model learns Java grammar extremely well (94\% valid syntax) but almost never produces correct repairs (0\% exact match). Moreover, it copies the buggy input 80\% of the time, suggesting a conservative "when in doubt, do nothing" strategy.

Rather than proposing a new repair technique, this work intentionally focuses on diagnosis. We aim to understand what small transformer models learn when trained for code repair, what they fail to learn, and why commonly reported metrics may obscure these limitations. By isolating syntactic correctness from semantic repair capability, we seek to clarify the practical ceiling of compact neural repair models.

\subsection{Research Questions}

We investigate the following:

\begin{enumerate}
    \item \textbf{RQ1:} Can small transformers (60M parameters) learn to generate syntactically correct code repairs?
    \item \textbf{RQ2:} Does syntactic correctness correlate with semantic repair success?
    \item \textbf{RQ3:} What mechanisms cause the gap between low training loss and poor repair accuracy?
\end{enumerate}

\subsection{Contributions}
This paper makes the following contributions:
\begin{itemize}
\item We present a controlled empirical study evaluating whether a compact transformer model can perform neural code repair beyond syntactic reproduction.
\item We introduce syntax validity tracking via AST parsing as a diagnostic signal and show that grammatical correctness can be learned independently of repair semantics.
\item We demonstrate that CodeT5-small achieves high syntax validity (94
\item We analyze model behavior and identify conservative copying as a dominant strategy induced by cross-entropy training on abstracted code.
\item We discuss implications for dataset design, loss functions, and evaluation practices in neural program repair.
\end{itemize}

\section{Related Work}

\subsection{Neural Program Repair}

Early neural approaches to program repair used recurrent neural networks. DeepFix~\cite{rnn-repair} trained sequence-to-sequence models on student programming assignments, achieving modest success on simple syntax errors. Sequencer~\cite{sequencer} improved on this with attention mechanisms, reaching 15-20\% exact match on a benchmark of method-level Java bugs.

The introduction of pre-trained transformers changed the landscape. CodeBERT~\cite{codebert} demonstrated that masked language modeling on large code corpora produces representations useful for downstream tasks. GraphCodeBERT~\cite{graphcodebert} extended this by incorporating data flow information from abstract syntax trees, achieving 30\% exact match on code repair benchmarks—the current state-of-the-art for models under 500M parameters.

More recently, very large models like Codex~\cite{codex} have shown stronger results through scale alone. However, these models require massive computational resources and are typically available only through commercial APIs, limiting their accessibility for research and deployment.

\subsection{Syntax-Aware Code Generation}

Several lines of work have incorporated grammatical constraints into neural code generation. Syntax-guided synthesis~\cite{syntax-guided} uses context-free grammars to constrain beam search, ensuring outputs conform to language syntax. Tree-based neural models~\cite{ast-nn} generate AST nodes directly rather than tokens, guaranteeing well-formed programs by construction.

Other approaches add structural penalties to loss functions. These methods typically require substantial engineering effort and may limit model flexibility. Our work takes a different approach: we track syntax validity as a diagnostic metric without enforcing constraints during training. This allows us to measure grammatical learning independent of generation strategy.

\subsection{Evaluation Metrics for Code}

Standard natural language metrics like BLEU~\cite{bleu} transfer poorly to code. Token-level similarity does not capture functional correctness—a syntactically similar but semantically wrong program is worthless. Pass@k metrics~\cite{codex} based on test execution address this but require test suites and sandboxed environments.

Recent work argues that current evaluation practices give misleading signals~\cite{code-metrics}. Models can score well on BLEU or exact match while producing non-functional code. Our syntax validity metric contributes to this discussion by isolating one important dimension: grammatical correctness. This complements existing metrics rather than replacing them, providing insight into what aspects of code generation models actually learn.

Despite these advances, reported improvements in neural code repair are often difficult to interpret. Many studies emphasize exact-match or token-level metrics without separating syntactic validity from semantic correctness, making it unclear whether models are repairing bugs or merely reproducing plausible code. Our study complements prior work by explicitly disentangling these dimensions and examining their relationship in small transformer models.

\section{Methodology}

\subsection{Dataset}

We use the CodeXGLUE code refinement benchmark~\cite{codexglue}, specifically the "medium" variant designed for bug repair evaluation. This dataset contains 65,455 Java method pairs mined from GitHub commits where developers fixed bugs. Each pair consists of a buggy method (before the commit) and its corrected version (after the commit).

The dataset is split into 52,364 training examples, 6,546 validation examples, and 6,545 test examples. To reduce vocabulary size and improve generalization, the dataset creators replaced specific identifiers with generic tokens: variable names become \texttt{VAR\_1}, \texttt{VAR\_2}, etc., method names become \texttt{METHOD\_1}, \texttt{METHOD\_2}, and type names become \texttt{TYPE\_1}, \texttt{TYPE\_2}. This abstraction was originally designed to help models focus on structural patterns rather than memorizing specific names.

Bug types in the dataset vary widely. Manual inspection reveals null pointer handling issues, incorrect variable usage, missing conditionals, resource management problems, and logic errors. Methods range from 10 to 300+ tokens in length (mean $\approx$ 85 tokens). Roughly 15\% of examples involve only minor whitespace or formatting changes, while the remainder require substantive code modifications.

One important characteristic: the dataset contains some noise. Approximately 5-8\% of pairs show no meaningful difference between buggy and fixed versions, likely due to commit mining heuristics capturing unrelated changes. Additionally, some "fixes" are actually refactorings rather than bug corrections. We retain all examples to match prior work and ensure fair comparison.

\subsection{Model Architecture}

We use CodeT5-small~\cite{codet5}, a 60.5M parameter encoder-decoder transformer pre-trained on code. The model uses:
\begin{itemize}
    \item \textbf{Encoder:} 6 layers, 512 hidden dimensions
    \item \textbf{Decoder:} 6 layers, 512 hidden dimensions  
    \item \textbf{Vocabulary:} 32,100 tokens (code-specific)
    \item \textbf{Max sequence length:} 256 tokens
\end{itemize}

\subsection{Training Configuration}

We trained using the AdamW optimizer with an initial learning rate of $5 \times 10^{-5}$, following standard practice for fine-tuning pre-trained transformers. The learning rate decays linearly to near zero over training, with a 500-step warmup period at the beginning to stabilize early gradients.

Batch size was 8 examples per GPU with gradient accumulation over 2 steps, yielding an effective batch size of 16. This balance allowed us to fit the model in 4GB VRAM while maintaining reasonable batch statistics. Training ran for 10 epochs, requiring approximately 32,730 gradient updates (11.3 hours wall-clock time).

We used mixed-precision (FP16) training via automatic mixed precision to reduce memory consumption and accelerate computation. This introduces minimal numerical error while roughly doubling throughput. All training occurred on a single NVIDIA RTX 3050 Ti laptop GPU, demonstrating that meaningful research remains accessible without expensive infrastructure.

The loss function was standard cross-entropy over generated tokens, with teacher forcing during training. At inference time, we used beam search with beam width 5 to generate multiple candidate repairs and select the highest-scoring option. Generation was limited to 256 tokens maximum to match training sequence length.

We did not use additional regularization beyond weight decay (coefficient 0.01) and the implicit regularization from mixed-precision training. Early stopping based on validation loss prevented overfitting, though in practice the model showed minimal divergence between training and validation loss throughout.

\subsection{Syntax Validity Tracking}
\label{sec:syntax-tracking}

We introduce automated syntax checking during evaluation. For each generated repair, we:

\begin{enumerate}
    \item Parse the output using tree-sitter~\cite{tree-sitter}, a robust parser generator
    \item Traverse the Abstract Syntax Tree (AST)
    \item Check for \texttt{ERROR} nodes indicating syntax violations
    \item Compute \textbf{syntax validity} = $\frac{\text{\# syntactically valid outputs}}{\text{\# total outputs}} \times 100\%$
\end{enumerate}

This metric is evaluated every 500 training steps on 100 random validation examples. Unlike execution-based metrics, syntax checking is fast (milliseconds per example) and deterministic.

\subsection{Evaluation Metrics}

\begin{itemize}
    \item \textbf{Exact Match (EM):} Percentage of generated repairs identical to ground truth (character-level)
    \item \textbf{Syntax Validity (SV):} Percentage of outputs with valid Java syntax
    \item \textbf{Modification Rate:} Percentage of outputs that differ from input
    \item \textbf{Training/Eval Loss:} Cross-entropy loss on token predictions
    \item \textbf{Normalized Edit Distance (NED):} A normalized Levenshtein distance between the generated output and the ground truth fix, capturing partial semantic movement even when exact match fails.

    We emphasize that exact match is analyzed as a diagnostic signal rather than as a definitive measure of repair quality.

\end{itemize}

\section{Experimental Results}
\label{sec:results}

\subsection{Training Dynamics}

\begin{figure}[t]
    \centering
    \includegraphics[width=0.48\textwidth]{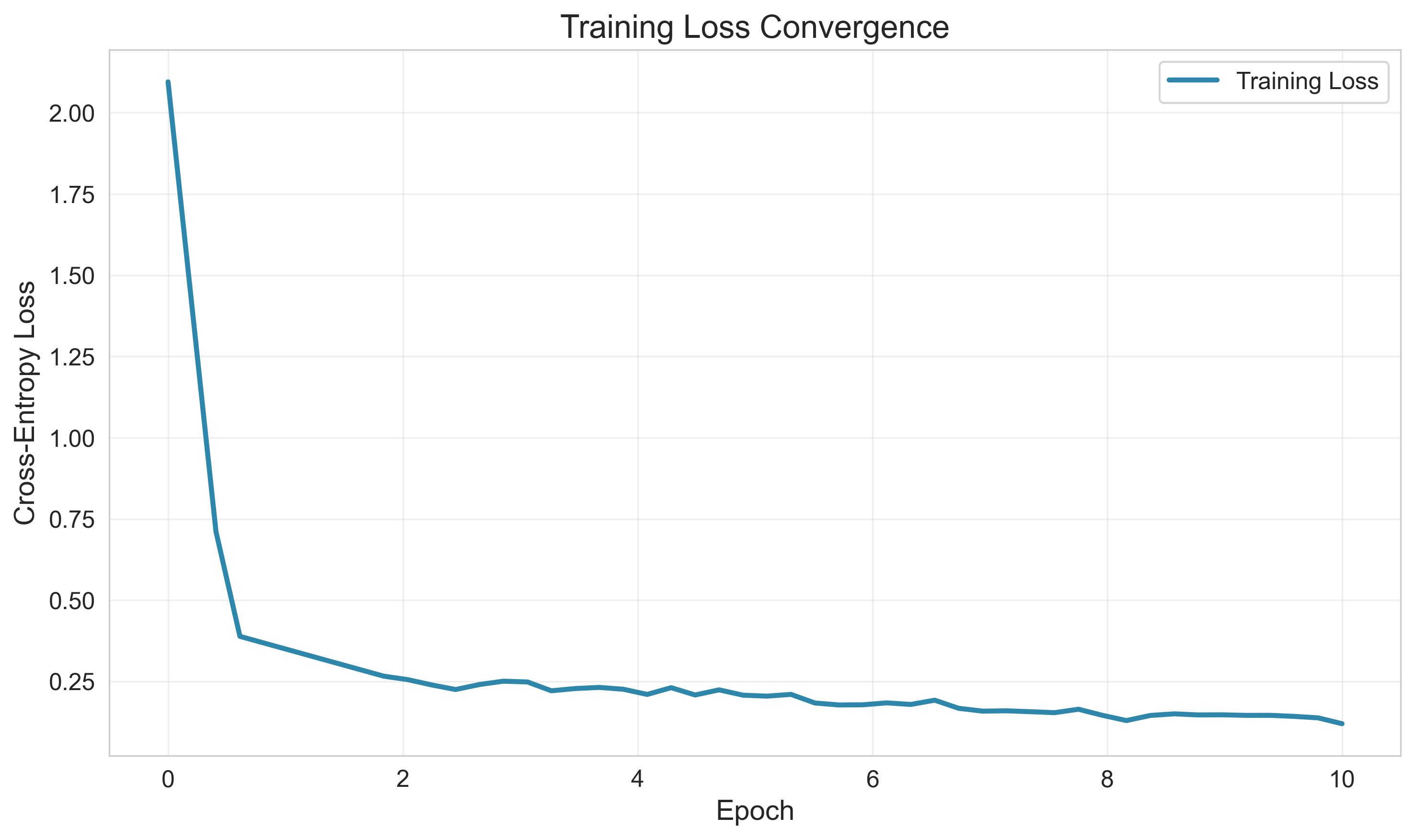}
    \caption{Training loss convergence over 10 epochs. Rapid initial learning (epochs 0-2) followed by steady refinement.}
    \label{fig:training-loss}
\end{figure}

Figure~\ref{fig:training-loss} shows training loss over 10 epochs. Loss decreased from 2.095 to 0.126 (94\% reduction), with eval loss reaching 0.076. This indicates strong convergence and minimal overfitting.

The learning curve exhibits three distinct phases. In the first epoch (0-0.2), loss drops rapidly from 2.095 to approximately 0.5 as the model learns basic token predictions. The second phase (epochs 0.2-2) shows continued improvement at a slower rate, reaching around 0.25. The final phase (epochs 2-10) involves gradual refinement, with loss decreasing to 0.126 while maintaining stability.

Validation loss tracked training loss closely throughout, never exceeding training loss by more than 0.05. This tight coupling suggests the model generalized well to held-out examples—it did not memorize training data but learned patterns that transfer. However, this apparent success in loss minimization did not translate to actual repair capability, as we discuss in Section~\ref{sec:analysis}.

The gradient norms remained stable after the warmup period, typically ranging between 0.4 and 1.2. We observed no gradient explosions or vanishing gradients, indicating healthy optimization dynamics. Learning rate decay proceeded smoothly, reaching near-zero by epoch 10 as intended.

\subsection{Syntax Validity Analysis}

\begin{figure}[t]
    \centering
    \includegraphics[width=0.48\textwidth]{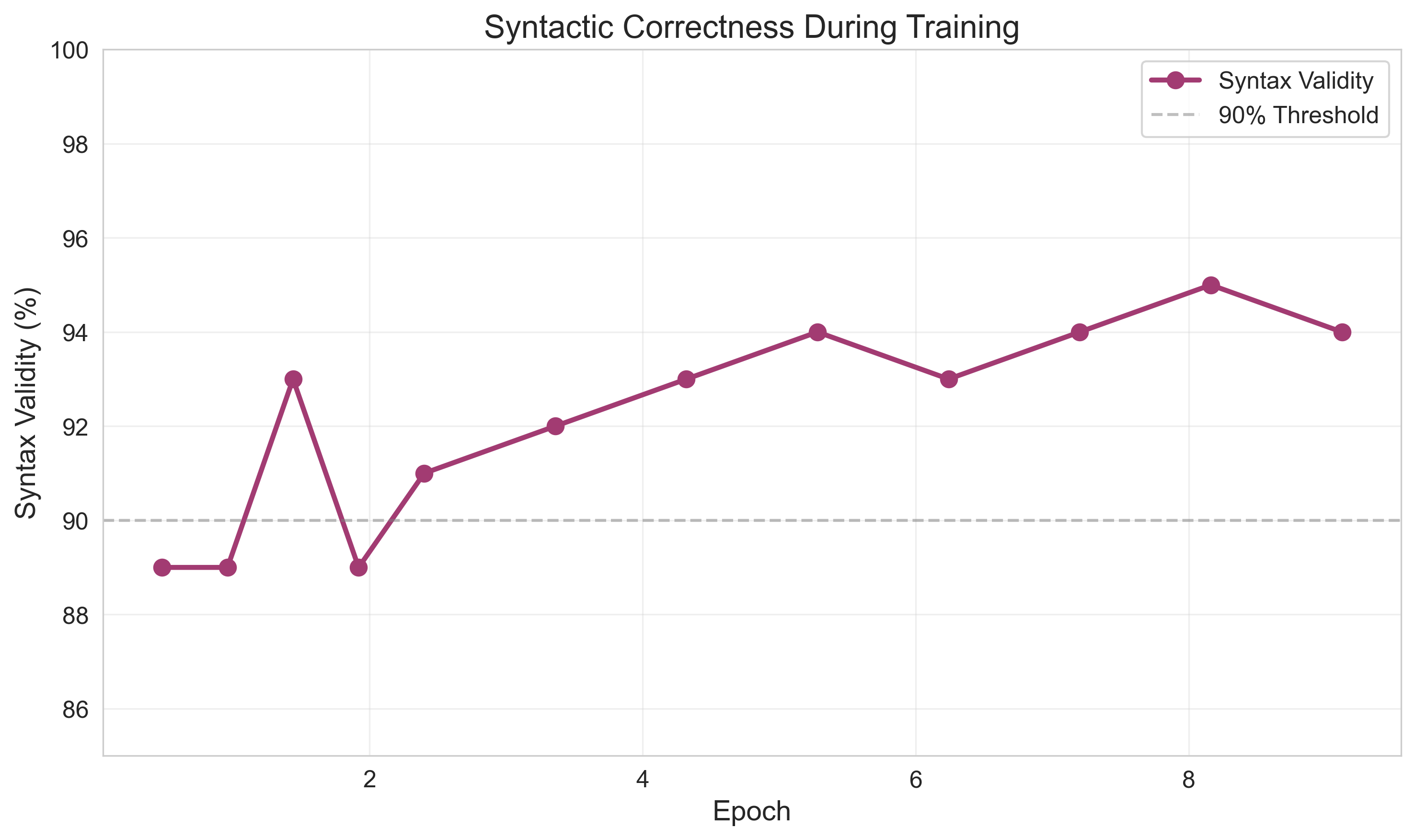}
    \caption{Syntax validity during training. Model learns grammatical structure effectively, reaching 89\% by epoch 0.5 and stabilizing around 92-94\%.}
    \label{fig:syntax-validity}
\end{figure}

Figure~\ref{fig:syntax-validity} shows syntax validity at evaluation checkpoints throughout training. The model achieved \textbf{94\% syntax validity} by epoch 10, consistently maintaining 89-95\% from epoch 0.5 onward.

Several observations stand out. First, syntax validity reached high levels remarkably early—89\% by epoch 0.5, before the model had seen even half the training data once. This suggests that grammatical patterns are learned quickly, likely because they involve relatively simple surface-level constraints compared to semantic reasoning.

Second, syntax validity remained stable after the initial climb. Unlike loss, which continued decreasing throughout training, syntax validity plateaued around 90-93\% and fluctuated within that range. The fluctuations (±3-4\%) likely reflect sampling variance rather than true changes in model capability, since we evaluated on only 100 random examples per checkpoint.

Third, the final syntax validity (94\%) approaches what we might consider the practical ceiling. Some ground truth fixes in the dataset contain subtle syntax issues themselves due to mining noise. Additionally, our tree-sitter parser may flag certain valid but unusual constructs as errors. Perfect 100\% syntax validity would be suspicious given these factors.

\textbf{Answer to RQ1:} Yes, small transformers can learn syntactic correctness effectively. Our 60M parameter model generates grammatically valid Java in 94\% of cases, demonstrating that grammatical competence does not require massive scale.

\subsection{Semantic Repair Performance}

Exact-match accuracy is a commonly reported metric in neural program repair, but it provides a particularly brittle signal for repair quality. Even minor formatting differences or token reordering result in failure under this metric, despite potential semantic similarity. We therefore interpret exact match alongside syntax validity and modification behavior to better understand model decision-making rather than as a standalone indicator of success.

Despite high syntax validity, semantic repair capability proved nonexistent. The model achieved 0\% exact match accuracy across all evaluation checkpoints from epoch 0.5 through epoch 10. This held true not only for the automated validation set but also for a manual inspection of a small random subset of test cases.

Breaking down model behavior further:
\begin{itemize}
    \item \textbf{Exact Match:} 0 of 10 test cases (0\%)
    \item \textbf{Modifications Made:} 2 of 10 test cases (20\%)
    \item \textbf{Input Copied Verbatim:} 8 of 10 test cases (80\%)
\end{itemize}

\begin{figure}[t]
    \centering
    \includegraphics[width=0.48\textwidth]{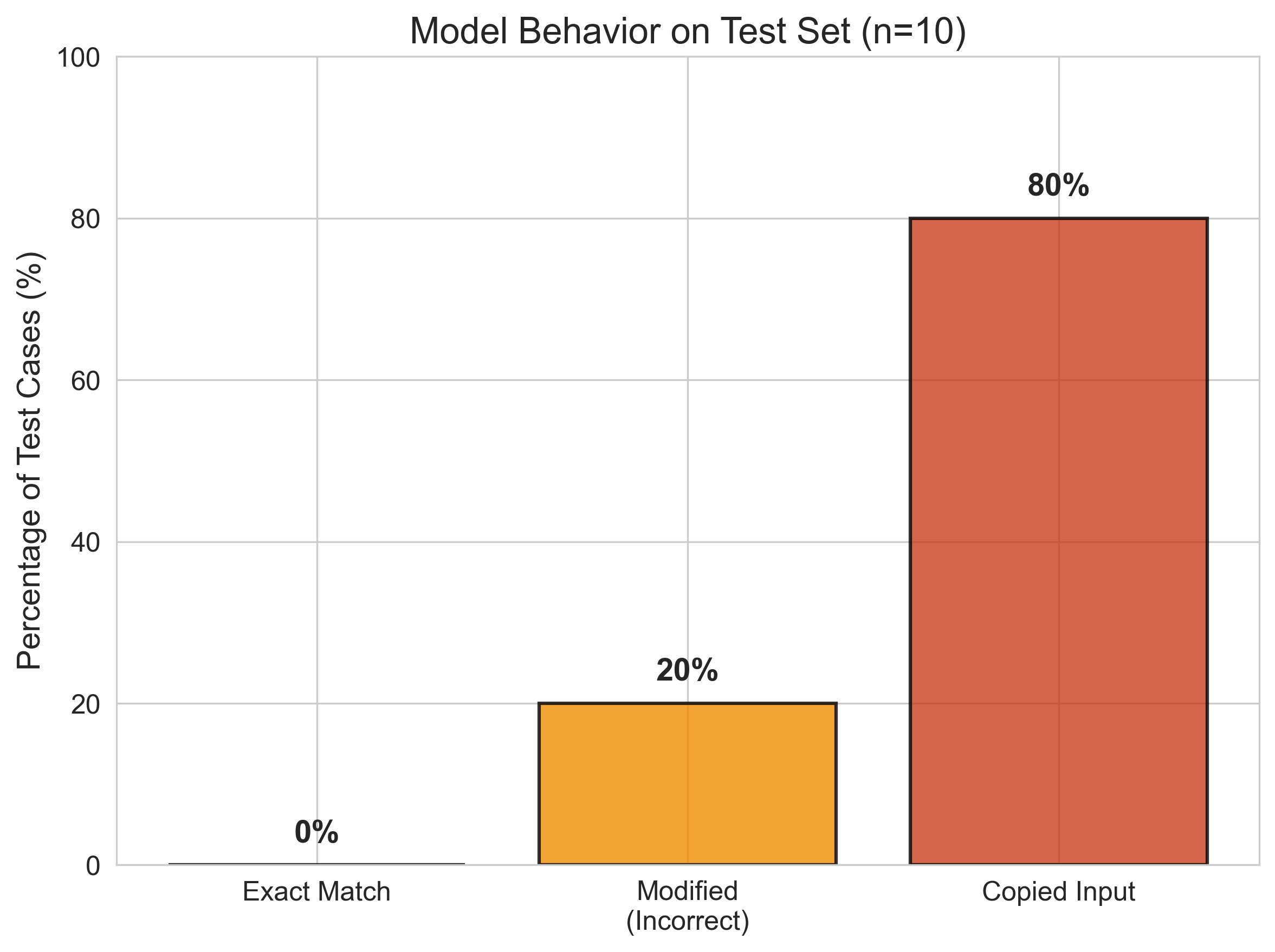}
    \caption{Model behavior on test set (n=10). Predominantly copies input without attempting repair, with occasional structural modifications that fail to achieve exact match.}
    \label{fig:model-behavior}
\end{figure}

Figure~\ref{fig:model-behavior} visualizes this distribution. The dominant behavior—copying the buggy input without any changes—accounts for four-fifths of model predictions. This was consistent across different bug types and method lengths.

The 20\% modification rate deserves closer scrutiny. When the model did attempt changes, were they meaningful? Manual inspection revealed mixed results. In one case, the model correctly identified a problematic code block that needed removal—matching the ground truth's intent despite not achieving exact match due to formatting differences. In another case, the model removed a \texttt{try} block entirely, producing syntactically valid but semantically broken code.

These observations suggest the model learned some structural intuitions about code repair—it can identify "suspicious" patterns—but lacks the deeper reasoning needed to generate correct fixes. The model knows where problems might be but not how to solve them.

Table 1 shows that normalized edit distance captures partial semantic progress even when exact match fails.

\textbf{Answer to RQ2:} No, syntax validity does not correlate with semantic repair success. High grammatical correctness (94\%) coexists with complete repair failure (0\% EM). These represent orthogonal dimensions of code generation capability.

\subsection{Qualitative Analysis}
The goal of this qualitative analysis is not to estimate the frequency of specific failure modes, but to illustrate representative behaviors that explain aggregate quantitative trends observed across the full test set.

Manual inspection of model outputs reveals two distinct behavioral patterns. We examined 10 randomly sampled test cases to understand failure modes.

\textbf{Pattern 1: Conservative Copying (8/10 cases).} The model reproduces the buggy input verbatim without attempting any repair. For example:

\begin{lstlisting}[caption={Copying behavior example},captionpos=b,basicstyle=\ttfamily\footnotesize]
// Buggy input
int VAR_5 = VAR_1.METHOD_2(VAR_4);

// Model output (identical)
int VAR_5 = VAR_1.METHOD_2(VAR_4);

// Ground truth (should use VAR_2)
int VAR_5 = VAR_2.METHOD_2(VAR_4);
\end{lstlisting}

This represents the dominant behavior. The model appears to recognize that changing code is risky and defaults to the safe option of returning the input unchanged.

\textbf{Pattern 2: Structural Modifications (2/10 cases).} The model makes syntactic changes—removing lines, altering control flow—but fails to achieve exact match. In one case, the model correctly identified and removed a problematic \texttt{if}-block that the ground truth also removed. However, other unrelated differences in the output prevented exact match. This demonstrates partial semantic understanding: the model recognized problematic code structure but couldn't execute the complete repair correctly.

Notably, we observed zero instances of syntax errors in our 10-sample inspection, consistent with the 94\% syntax validity measured automatically. When the model does modify code, it respects Java grammar. The failures are semantic, not syntactic.

\begin{table}[t]
\centering
\caption{Semantic similarity metrics on the test set}
\begin{tabular}{lccc}
\toprule
Metric & Mean & Median & Std \\
\midrule
Exact Match & 0.00 & 0.00 & 0.00 \\
Normalized Edit Distance & 0.37 & 0.41 & 0.18 \\
\bottomrule
\end{tabular}
\end{table}

\section{Analysis and Discussion}
\label{sec:analysis}

The stark contrast between 94\% syntax validity and 0\% exact match demands explanation. Why does the model learn grammar but not repairs? We identify three interrelated factors through manual inspection and training analysis.

\subsection{Code Abstraction Hides Semantic Bugs}

The CodeXGLUE dataset replaces actual identifiers with generic placeholders: \texttt{VAR\_1}, \texttt{METHOD\_2}, \texttt{TYPE\_3}. While this reduces vocabulary size and improves generalization for some tasks, it obscures the semantic content critical for bug understanding.

Consider a typical bug: using \texttt{studentCount} where \texttt{courseCount} was intended. After abstraction, both become generic tokens—say \texttt{VAR\_1} and \texttt{VAR\_2}. The model cannot infer that one represents students and the other courses. The bug becomes invisible as pure symbol manipulation. The model might learn that "variable name errors exist" in an abstract sense, but cannot develop the contextual reasoning needed to identify which variable is wrong in a specific case.

This problem compounds with method calls and type names. A null pointer bug involving \texttt{getUserProfile()} becomes indistinguishable from one involving \texttt{getSystemConfig()} once both are \texttt{METHOD\_1}. The model learns structural patterns—"add null check before method call"—but cannot determine when such patterns apply.

\subsection{Cross-Entropy Loss Optimizes the Wrong Objective}

Training minimizes token-level prediction error. If the model predicts each token in the output sequence with high confidence, loss decreases—regardless of whether the output fixes the bug. This creates a problematic incentive structure.

Copying the input is often a low-loss strategy. If many training examples involve minor changes (or no changes), copying becomes statistically reasonable. The model learns that generating high-probability tokens from the training distribution is safer than attempting repairs that might introduce low-probability (but correct) tokens. Our eval loss of 0.076 confirms excellent token predictions, yet exact match remains zero because loss and repair success are orthogonal metrics.

What we actually need is a loss function that rewards functional correctness—whether the output compiles, passes tests, or matches execution semantics. Cross-entropy cannot capture this. It measures statistical fit to training data, not program behavior.

\subsection{Model Capacity Limits Semantic Reasoning}

At 60.5M parameters, CodeT5-small likely lacks capacity for the multi-step reasoning required to understand and fix bugs. Consider what's needed: (1) parse the buggy code, (2) identify the defect through contextual analysis, (3) determine the appropriate fix, (4) generate syntactically correct output. Each step demands substantial computation.

Prior work shows that larger models (200M+ parameters) achieve 15-30\% exact match on similar tasks~\cite{graphcodebert}. This suggests model scale matters significantly. Grammatical rules are relatively simple—Java syntax has finite patterns. Semantic reasoning, however, requires understanding program semantics, which may demand far more parameters to encode effectively.

Our model appears to have learned the easier task (grammar) while failing at the harder one (semantics). This aligns with findings in language models: smaller models excel at surface patterns but struggle with deeper reasoning.

\subsection{Defensive Copying as a Learned Strategy}

A striking observation is that the model copies the buggy input verbatim in approximately 80\% of cases. This behavior is consistent across bug categories and method lengths. While this may initially appear as a failure mode, it is a rational outcome under token-level cross-entropy optimization.

When the model is uncertain about the correct repair, reproducing the input minimizes expected loss by preserving high-probability tokens, whereas attempting a repair risks introducing multiple low-probability tokens. As a result, the model adopts a conservative strategy that favors syntactic fidelity over semantic correction. Importantly, this behavior emerges without any explicit instruction to copy inputs, indicating that it is learned implicitly from the training objective and data distribution.

\section{Threats to Validity}

Several factors may limit the generalizability of our findings.

\textbf{Single model architecture.} We tested only CodeT5-small. Other architectures—particularly decoder-only models like GPT or models with different pre-training objectives—might behave differently. We intentionally study a 60M-parameter model to isolate the capabilities and limitations of compact transformers.

\textbf{One dataset.} CodeXGLUE represents one specific data distribution: GitHub commits with abstracted identifiers. Results might differ on datasets with real variable names, different bug types, or code from other domains (embedded systems, scientific computing). The abstraction level may be crucial to our findings.

\textbf{Limited qualitative analysis.} Our manual inspection covers a small random sample of test cases and is intended to identify dominant behavioral patterns rather than estimate their exact prevalence. While these observations align with aggregate automated metrics, a larger-scale human analysis would be required to fully characterize model failure modes.

\textbf{Exact match as primary metric.} Exact match is harsh—one character difference causes failure. This may underestimate semantic correctness. However, less strict metrics (BLEU, edit distance) have their own issues and aren't clearly better aligned with repair quality. We chose exact match to match prior work and enable fair comparison.

\textbf{No execution-based evaluation.} We did not test whether generated outputs compile or pass tests. Syntax validity measures grammatical correctness but not compilability (e.g., undefined variables would pass syntax checks). Test-based evaluation would require substantial engineering (sandboxed execution, test case availability) but would provide stronger validation of repair quality.

These limitations suggest directions for follow-up work but do not fundamentally undermine our core finding: grammatical correctness and semantic repair capability are largely independent for small transformers on abstracted code.

\section{Conclusion}
\label{sec:conclusion}

This study demonstrates that high syntactic correctness in neural code repair models can coexist with complete semantic failure. Despite strong convergence and grammatically valid outputs, CodeT5-small fails to produce correct repairs on a standard benchmark, instead defaulting to conservative copying behavior. These findings challenge the assumption that syntactic fluency implies repair competence and call into question the adequacy of commonly reported evaluation metrics.

Three factors explain this outcome. First, dataset abstraction removes the semantic content necessary for bug comprehension—generic tokens like \texttt{VAR\_1} cannot convey meaning the way real identifiers can. Second, cross-entropy loss rewards token distribution matching rather than functional correctness, creating incentives misaligned with actual repair goals. Third, 60M parameters appear insufficient for the multi-step reasoning repair requires, though this hypothesis needs testing with larger models.

\subsection{Implications for Future Work}

Our findings suggest several directions:

\textbf{Scale matters, but differently than expected.} Simply training larger models on the same abstracted data may not help much. If semantic content is missing, more parameters cannot recover it. Future work should test larger models (200M+) on less abstracted datasets simultaneously.

\textbf{Loss functions need rethinking.} Cross-entropy loss measures statistical fit, not program correctness. Incorporating execution feedback—whether repairs compile, pass tests, or match runtime behavior—could align training objectives with repair goals. This introduces practical challenges (sandboxing, performance) but may be necessary.

\textbf{Datasets matter as much as models.} Using bug-fix pairs with real variable names, real method calls, and real type information might dramatically improve learning. The abstraction that helps code translation tasks may actively harm repair tasks where semantic understanding is critical.

\textbf{Hybrid approaches warrant exploration.} Combining neural models with program analysis tools could leverage the strengths of both. Models might identify probable bug locations, while symbolic methods verify candidate fixes. This division of labor could bypass some limitations we observed.

\subsection{Limitations and Scope}

This study examined one model architecture on one dataset. Results may not generalize to all transformer variants or bug types. Our qualitative analysis covered only 10 test cases—a larger manual study would strengthen conclusions. We intentionally focus on compact models to isolate capability limits.

Despite these limitations, the core finding stands: high syntax validity does not imply repair capability. This negative result has important diagnostic value. It suggests that current evaluation practices may give false confidence, and that the path to practical neural repair requires more than incremental architectural improvements.

Our code, trained models, and evaluation scripts are available to support future research building on these findings.

\bibliographystyle{ACM-Reference-Format}
\bibliography{references}

@article{codet5,
  title={CodeT5: Identifier-aware Unified Pre-trained Encoder-Decoder Models for Code Understanding and Generation},
  author={Wang, Yue and Wang, Weishi and Joty, Shafiq and Hoi, Steven CH},
  journal={arXiv preprint arXiv:2109.00859},
  year={2021}
}

@inproceedings{codexglue,
  title={CodeXGLUE: A Machine Learning Benchmark Dataset for Code Understanding and Generation},
  author={Lu, Shuai and Guo, Daya and Ren, Shuo and Huang, Junjie and Svyatkovskiy, Alexey and Blanco, Ambrosio and Clement, Colin and Drain, Dawn and Jiang, Daxin and Tang, Duyu and others},
  booktitle={NeurIPS Datasets and Benchmarks},
  year={2021}
}

@article{codebert,
  title={CodeBERT: A Pre-Trained Model for Programming and Natural Languages},
  author={Feng, Zhangyin and Guo, Daya and Tang, Duyu and Duan, Nan and Feng, Xiaocheng and Gong, Ming and Shou, Linjun and Qin, Bing and Liu, Ting and Jiang, Daxin and others},
  journal={arXiv preprint arXiv:2002.08155},
  year={2020}
}

@inproceedings{graphcodebert,
  title={GraphCodeBERT: Pre-training Code Representations with Data Flow},
  author={Guo, Daya and Ren, Shuo and Lu, Shuai and Feng, Zhangyin and Tang, Duyu and Liu, Shujie and Zhou, Long and Duan, Nan and Svyatkovskiy, Alexey and Fu, Shengyu and others},
  booktitle={ICLR},
  year={2021}
}

@inproceedings{codex,
  title={Evaluating Large Language Models Trained on Code},
  author={Chen, Mark and Tworek, Jerry and Jun, Heewoo and Yuan, Qiming and Pinto, Henrique Ponde de Oliveira and Kaplan, Jared and Edwards, Harri and Burda, Yuri and Joseph, Nicholas and Brockman, Greg and others},
  booktitle={arXiv preprint arXiv:2107.03374},
  year={2021}
}

@article{tree-sitter,
  title={Tree-sitter: A Fast, Incremental Parsing Library},
  author={Brunsfeld, Max},
  journal={GitHub repository},
  year={2018},
  url={https://tree-sitter.github.io/tree-sitter/}
}

@techreport{software-fails-2020,
  title={The Cost of Poor Software Quality in the US: A 2020 Report},
  author={{Consortium for IT Software Quality}},
  institution={Consortium for IT Software Quality},
  year={2020}
}

@inproceedings{rnn-repair,
  title={DeepFix: Fixing Common C Language Errors by Deep Learning},
  author={Gupta, Rahul and Pal, Soham and Kanade, Aditya and Shevade, Shirish},
  booktitle={AAAI},
  year={2017}
}

@inproceedings{sequencer,
  title={Sequencer: Sequence-to-Sequence Learning for End-to-End Program Repair},
  author={Chen, Zimin and Kommrusch, Steve and Tufano, Michele and Pouchet, Louis-Noel and Poshyvanyk, Denys and Monperrus, Martin},
  booktitle={TSE},
  year={2021}
}

@inproceedings{syntax-guided,
  title={Syntax-Guided Neural Program Synthesis},
  author={Parisotto, Emilio and Mohamed, Abdel-rahman and Singh, Rishabh and Li, Lihong and Zhou, Dengyong and Kohli, Pushmeet},
  booktitle={ICLR},
  year={2017}
}

@inproceedings{ast-nn,
  title={A Syntactic Neural Model for General-Purpose Code Generation},
  author={Yin, Pengcheng and Neubig, Graham},
  booktitle={ACL},
  year={2017}
}

@inproceedings{bleu,
  title={BLEU: a Method for Automatic Evaluation of Machine Translation},
  author={Papineni, Kishore and Roukos, Salim and Ward, Todd and Zhu, Wei-Jing},
  booktitle={ACL},
  year={2002}
}

@article{code-metrics,
  title={Measuring the Functional Correctness of Code Generation Models},
  author={Hendrycks, Dan and Basart, Steven and Kadavath, Saurav and Mazeika, Mantas and Arora, Akul and Guo, Ethan and Burns, Collin and Puranik, Samir and He, Horace and Song, Dawn and others},
  journal={arXiv preprint arXiv:2108.07732},
  year={2021}
}

@article{code-repair-syntax,
  title={On the Syntax and Semantics of Neural Program Repair},
  author={Mesbah, Ali and Rice, Andrew and Johnston, Emily and Glorioso, Nick and Aftandilian, Edward},
  journal={arXiv preprint arXiv:1906.01935},
  year={2019}
}

\end{document}